\documentstyle[aps]{revtex}

\title{Quantum dynamics in canonical and micro-canonical ensembles.
Part I. Anderson localization of electrons.}

\author{
V.S. Filinov\thanks{Other author information: (Send correspondence to V.S.F.)
V.S.F.: Email: filinov@vovan.msk.ru; Telephone: 7(095)931-07-19;
Fax: 7(095)485-79-90}, \ \\
{\it Russian Academy of Sciences, 'IVTAN' Association \ \ \\
High Energy Density Research Center \ \ \\
Izhorskaya str. 13/19, Moscow,
\hspace{0.5em}127412, \hspace{0.5em}Russia} \\
Yu. E. Lozovik, A. V. Filinov \ \\
{\it Russian Academy of Sciences, Institute of Spectroscopy,  \ \ \\
\hspace{0.5em}Troitsk, Moscow region, 142092, \hspace{0.5em}Russia} \ \\
I. Zacharov \ \ \\
{\it Silicon Graphics Computer Systems, SGI Europe, \ \ \\
Grand Atrium, route des Avouillons 30, \\
\hspace{0.5em}1196  Gland, \hspace{0.5em}Switzerland}  \ \\
Alexei M. Oparin \\
{\it Russian Academy of Sciences, Institute for Computer Aided Design \ \\
Vtoraya Brestskaya str. 19/18, \ \\
\hspace{0.5em}Moscow, 123056, \hspace{0.5em}Russia}
}

\begin{document}
\maketitle
\hspace{3cm} Phys. Abstr. Class.: 72.10.Bg; 02.10.-c; 02.70-c; 02.50-r; 02.60-x
\begin{abstract}
The new numerical approach for consideration of quantum dynamics and
calculations of the average values of quantum operators and time correlation
functions in the Wigner representation of quantum statistical mechanics has
been developed. The time correlation functions have been presented in the
form of the integral of the Weyl's symbol of considered operators and the
Fourier transform of the product of matrix elements of the dynamic
propagators. For the last function the integral Wigner- Liouville's type
equation has been derived. The initial condition for this equation has been
obtained in the form of the Fourier transform of the Wiener path integral
representation of the matrix elements of the propagators at initial time.
The numerical procedure for solving this equation combining both molecular
dynamics and Monte Carlo methods has been developed. An application of the
developed approach to the micro canonical ensemble has been also considered
in the second part of this paper.

For electrons in disordered systems of scatterers the numerical results have
been obtained for series of the average values of the quantum operators
including position and momentum dispersions, average energy, energy
distribution function as well as for the frequency dependencies of tensor of
electron conductivity and permittivity according to quantum Kubo formula.
Zero or very small value of static conductivity have been considered as the
manifestation of Anderson localization of electrons in 1D case. Independent
evidence of Anderson localization comes from the behaviour of the calculated
time dependence of position dispersion. Nevertheless for localized electrons
the energy distribution function obtained has the long exponentially
decaying tail, which is the reason of exponentially rare appearance of large
values of quantum particle virtual energy that strongly affects the behaviour
of the position dispersion.
\end{abstract}

\date{July 1997}


\newpage




\section{Introduction}

It is known that crystalline materials have been studied intensively by
physicists for a long time since the beginning of quantum physics.
Practically the same is valid for investigations of the disorder state of
the matter. Disorder exist in varying degree, ranging from the a few
impurities or interstitials in an otherwise perfect crystalline host to the
strongly disordered limit of alloys or glassy structures. In the past few
years there has been growing realization that for understanding the
disordered materials new concepts must be introduced which treat the
disorder from the very beginning \cite{leeram}. The new understanding is
based on advances in two different areas . The first is the problem of
Anderson localization, which deals with the nature of the wave function of
single electron in the presence of the random potential. A scaling
description of the Anderson localization problem has greatly deepened
understanding of the problem. The second aspect of the problem is the
interaction among electrons in the presence of a random potential. The fact
that the electrons are diffusive instead of freely propagating leads to a
profound modification of the traditional view based on the Fermi-liquid
theory of metals.

In this paper we treat the first of the mentioned above problem. In 1958,
Anderson pointed out that if disorder is strong enough, the electron wave
function may become localized , in that the its envelope decays
exponentially from the initial electron position in space. The
characteristic space scale of this decay is called as localization length.
The energy and density dependence of localization length results in the metal-
insulator
transition and the concepts of the mobility energetic edge. Despite of
significant advances the mentioned and a lot of other related problem are
unsolved now due to thire complexity. For investigation of the problem of
Anderson localization we have applied the developed in \cite{filmd1} , \cite
{filmd2} numerical approach allowing rigorous calculations of the quantum
time correlation functions and average values of quantum operators in the
Wigner formulation of quantum statistical mechanics.

Within Wigner formulation of quantum mechanics the time correlation function
can be presented in the form of the integral of Weil's symbols of the
operators $\hat{F}$ and $\hat{A}$ and so called spectral density, which is
the Fourier transform of the product of the dynamic propagator matrix
elements arising due to Heisenberg representation of the operator $\hat{A}%
(t) $. For spectral density the integral equation of the Wigner- Liouville'
type has been derived. The developed approach for solving this equation
combines both molecular dynamics and Monte Carlo methods and allows to
combine the existing molecular dynamics and Monte Carlo codes after small
modernization. Efficiency of the developed method is high enough to use for
calculations not very fast computers. The explicit expression of the initial
condition for this integral equation is convenient to obtain as Fourier
transform of the propagator matrix elements presented in the form of Wiener
path integrals.

The numerical results have been obtained for series of the average value of
the quantum operators as well as for the frequency dependencies of tensor of
electron conductivity and permittivity according to quantum Kubo formulas
for electrons in disordered systems of scatterers. Zero or very small value
of static conductivity have been obtained as the manifestation of Anderson
localization of electrons in 1D case. Independent
evidence of Anderson localization comes from the behaviour of the calculated
time dependence of position dispersion. Nevertheless for localized electrons
the energy distribution function obtained has the long exponentially
decaying tail, which is the reason of exponentially rare appearance of large
values of quantum particle virtual energy that strongly affects the behaviour
of the position dispersion.

\section{Wigner representation of quantum statistical mechanics}

The time correlation functions $C_{FA}(t)=\left\langle \hat{F}(0)\hat{A}%
(t)\right\rangle $ for different dynamic properties are among the most
interesting quantities in studying dynamics of electrons in disordered
systems of scatters, transport coefficients, chemical reaction rates,
consideration of equilibrium and transient spectroscopy and so on. Our
starting point is the general operator expression for canonical ensemble
averaged time correlation functions \cite{zubar} :
\[
C_{FA}\left( t\right) =Z^{-1}Tr\left( \hat{F}\exp \left( i\hat{H}%
t_c^{*}/\hbar \right) \hat{A}\exp \left( -i\hat{H}t_c/\hbar \right) \right)
\]
Here $\hat{H}$ is the Hamiltonian of the system, $\hat{H}=\hat{K}+\hat{U}$, $%
\hat{K}$ is the kinetic energy operator, $\hat{U}$ is the potential energy
operator, $t_c=t-i\hbar \beta /2$, $\beta =1/k_BT$, $\hat{F}$ and $\hat{A}$
are quantum operators of the considered dynamic quantities, $Z=Tr\left( \exp
\left( -\beta \hat{H}\right) \right) $ is the partition function. Wigner
representation of the time correlation function can be written as:
\[
\begin{array}{c}
C_{FA}\left( t\right) =\frac 1{\left( 2\pi \hbar \right) ^{2\upsilon }}\int
\int dp_1dq_1dp_2dq_2F\left( p_1,q_1\right) A\left( p_{2,}q_2\right) \times
\\
\times W\left( p_{1,}q_1;p_{2,}q_2;t;i\hbar \beta \right)
\end{array}
\]
where the spectral density $W\left( p_{1,}q_1;p_{2,}q_2;t;i\hbar \beta
\right) $ is defined by:
\[
\begin{array}{c}
W\left( p_{1,}q_1;p_{2,}q_2;t;i\hbar \beta \right) =Z^{-1}\int \int d\xi
_1d\xi _2\exp \left( i\frac{p_1\xi _1}\hbar \right) \exp \left( i\frac{%
p_2\xi _2}\hbar \right) \times \\
\times \left\langle q_1+\frac{\xi _1}2\left| \exp \left( i\hat{H}%
t_c^{*}/\hbar \right) \right| q_2-\frac{\xi _2}2\right\rangle \left\langle
q_2+\frac{\xi _2}2\left| \exp \left( -i\hat{H}t_c/\hbar \right) \right| q_1-%
\frac{\xi _1}2\right\rangle
\end{array}
\]
and $F\left( p_1,q_1\right) $ and $A\left( p_{2,}q_2\right) $ are Weyl's
symbols of operators \cite{tatr1} $\hat{F}$ and $\hat{A}$ :
\[
A\left( p,q\right) =\int d\xi \exp \left( -i\frac{p\xi }\hbar \right)
\left\langle q-\frac \xi 2\left| \hat{A}\right| q+\frac \xi 2\right\rangle
\]
where $\left\langle q^{\prime }\left| \hat{A}\right| q^{\prime \prime
}\right\rangle $ are the matrix elements and $\upsilon $ is the space
dimension. So the problem of numerical calculation of the canonically
averaged time correlation function can be reduced to the consideration of
the spectral density evolution describing as can been proved by the
following integral equation :
\begin{equation}
\begin{array}{c}
W(p_1,q_1;p_2,q_2;t;i\hbar \beta )=\bar{W}(\bar{p}_0,\bar{q}_0;\tilde{p}_0,%
\tilde{q}_0;i\hbar \beta )+ \\
+\int_0^td\tau \int dsd\eta W(\bar{p}_\tau -s,\bar{q}_\tau ;\tilde{p}_\tau
-\eta ,\tilde{q}_\tau ;\tau ;i\hbar \beta )\gamma (s,\bar{q}_\tau ;\eta ,%
\tilde{q}_\tau )
\end{array}
\label{a22}
\end{equation}
where $\gamma $$(s,\bar{q}_\tau ;\eta ,\tilde{q}_\tau )=\frac 12\{\varpi
\left( s,\bar{q}_\tau \right) \delta (\eta )-\varpi \left( \eta ,\tilde{q}%
_\tau \right) \delta (s)\}$, $\delta (s)$ is Dirac delta function, $\varpi
\left( s,q\right) $ is defined by the expression
\[
\varpi \left( s,q\right) =\frac 4{(2\pi \hbar )^\upsilon \hbar }\int
dq^{\prime }U\left( q-q^{\prime }\right) \sin \left( \frac{2sq^{\prime }}%
\hbar \right) +\breve{F}\left( q\right) \frac{d\delta \left( s\right) }{ds}
\]
$\breve{F}\left( q\right) $ is the classical force, $\{\bar{q}_\tau (\tau
;p_1,q_1,t),$ $\bar{p}_\tau (\tau ;p_1,q_1,t)\}$ and $\left\{ \tilde{q}_\tau
(\tau ;p_2,q_2,t),\ \tilde{p}_\tau (\tau ;p_2,q_2,t)\right\} $ are the pare
of dynamic $pq$-trajectories for 'negative and positive time direction'
respectively and initial condition at $\tau =t$:
\begin{equation}
\begin{array}{c}
d\bar{p}/d\tau =\breve{F}(\bar{q}_\tau (\tau ))/2;\ \bar{q}%
_t(t;p_1,q_1,t)=q_1 \\
d\bar{q}/d\tau =\bar{p}_\tau (\tau )/2m;\ \ \bar{p}_t(t;p_1,q_1,t)=p_1
\end{array}
\label{a25}
\end{equation}
\[
\begin{array}{c}
d\tilde{p}/d\tau =-\breve{F}(\tilde{q}_\tau (\tau ))/2;\ \tilde{q}%
_t(t;p_2,q_2,t)=q_2 \\
d\tilde{q}/d\tau =-\tilde{p}_\tau (\tau )/2m;\ \tilde{p}_t(t;p_2,q_2,t)=p_2
\end{array}
\]
The spectral density initial condition $\bar{W}(p_{1,}q_1;p_{2,}q_2;i\hbar
\beta )\equiv W\left( p_{1,}q_1;p_{2,}q_{2,};0;i\hbar \beta \right) $ can be
written in the form of the finite difference approximation of the path
integral \cite{feynm} :
\[
\begin{array}{c}
\bar{W}\left( p_1,q_1;p_2,q_2;i\hbar \beta \right) \approx \\
\approx \int \int d\tilde{q}_{1...}d\tilde{q}_M\int \int dq_{1...}^{\prime
}dq_M^{\prime }\Psi \left( p_1,q_1;p_2,q_2;\tilde{q}_1,...,\tilde{q}%
_M;q_1^{\prime },...,q_M^{\prime };i\hbar \beta \right)
\end{array}
\]
\begin{equation}
\begin{array}{c}
\Psi \left( p_1,q_1;p_2,q_2;\tilde{q}_1,...,\tilde{q}_M;q_1^{\prime
},...,q_M^{\prime };i\hbar \beta \right) = \\
Z^{-1}\left\langle q_1\left| \exp \left( -\epsilon \hat{K}\right) \right|
\tilde{q}_1\right\rangle \exp \left( -\epsilon U\left( \tilde{q}_1\right)
\right) \left\langle \tilde{q}_1\left| \exp \left( -\epsilon \hat{K}\right)
\right| \tilde{q}_2\right\rangle \exp \left( -\epsilon U\left( \tilde{q}%
_2\right) \right) \times \\
...\exp \left( -\epsilon U\left( \tilde{q}_M\right) \right) \left\langle
\tilde{q}_M\left| \exp \left( -\epsilon \hat{K}\right) \right|
q_2\right\rangle \varphi \left( p_2;\tilde{q}_M,q_1^{\prime }\right) \times
\\
\left\langle q_2\left| \exp \left( -\epsilon \hat{K}\right) \right|
q_1^{\prime }\right\rangle \exp \left( -\epsilon U\left( q_1^{\prime
}\right) \right) \left\langle q_1^{\prime }\left| \exp \left( -\epsilon \hat{%
K}\right) \right| q_2^{\prime }\right\rangle \exp \left( -\epsilon U\left(
q_2^{\prime }\right) \right) \times \\
...\exp \left( -\epsilon U\left( q_M^{\prime }\right) \right) \left\langle
q_M^{\prime }\left| \exp \left( -\epsilon \hat{K}\right) \right|
q_1\right\rangle \varphi \left( p_1;q_M^{\prime },\tilde{q}_1\right) \\
\varphi \left( p;q^{\prime },q^{\prime \prime }\right) =\left( 2\lambda
^2\right) ^{\upsilon /2}\exp \left( -\frac{\left\langle p\lambda /\hbar
+i\pi \left( q^{\prime }-q^{\prime \prime }\right) /\lambda |\ p\lambda
/\hbar +i\pi \left( q^{\prime }-q^{\prime \prime }\right) /\lambda
\right\rangle }{2\pi }\right)
\end{array}
\label{a11}
\end{equation}
where $\epsilon =\frac \beta {2M}$, $M\gg 1$ and $\lambda ^2=2\pi \hbar
^2\beta /2mM$ .

Let us rewrite the integral equation (\ref{a22}) and the iteration form of
its solution in symbolic form: $W^t=\bar{W}^t+K_\tau ^tW^\tau $ and
\begin{equation}
W^t=\bar{W}^t+K_{\tau _1}^t\bar{W}^{\tau _1}+K_{\tau _2}^tK_{\tau _1}^{\tau
_2}\bar{W}^{\tau _1}+K_{\tau _3}^tK_{\tau _2}^{\tau _3}K_{\tau _1}^{\tau _2}%
\bar{W}^{\tau _1}+...  \label{r3}
\end{equation}
Here $\bar{W}^t$ and $\bar{W}^{\tau _1}$ is the quantum initial density
evolving classically in intervals $\left[ 0,t \right] $ and $\left[ 0,\tau
_1\right] $, while $K_{\tau _i}^{\tau _{i+1}}$ are operators, which describe
propagation between times $\tau _i$ and $\tau _{i+1}$. The time correlation
functions are the linear functionals of the spectral density:
\begin{equation}
\begin{array}{c}
C_{FA}\left( t\right) =\frac 1{\left( 2\pi \hbar \right) ^{2\upsilon }}\int
\int dp_1dq_1dp_2dq_2F\left( p_1,q_1\right) A\left( p_{2,}q_2\right) \times
\\
\times W\left( p_{1,}q_1;p_{2,}q_2;t;i\hbar \beta \right) = \\
\left( \phi |\bar{W}^t\right) +\left( \phi |K_{\tau _1}^t\bar{W}^{\tau
_1}\right) +\left( \phi |K_{\tau _2}^tK_{\tau _1}^{\tau _2}\bar{W}^{\tau
_1}\right) +\left( \phi |K_{\tau _3}^tK_{\tau _2}^{\tau _3}K_{\tau _1}^{\tau
_2}\bar{W}^{\tau _1}\right) +...
\end{array}
\label{a31}
\end{equation}
where brackets $\left( |\right) $ for functions $\phi \left(
p_{1,}q_1;p_{2,}q_2\right) =F\left( p_1,q_1\right) A\left( p_{2,}q_2\right) $
and $\bar{W}(\bar{p}_0,\bar{q}_0;\tilde{p}_0,\tilde{q}_0;i\hbar \beta )$ or $%
K_{\tau _i}^tK_{\tau _{i-1}}^{\tau _i}...K_{\tau _1}^{\tau _2}\bar{W}^{\tau
_1}$ mean the integration over the phase spaces $\left\{
p_{1,}q_1;p_{2,}q_2\right\} $.

Note that average values of quantum operators $\bar{F}\left( t\right) $ can
be formally presented in the form analogous to (\ref{a31}) :
\[
\bar{F}\left( t\right) =\tilde Z^{-1}Tr\left( \exp \left( i\hat{H}t/\hbar
\right) \hat{F}\exp \left( -i\hat{H}t/\hbar \right) \rho \left( 0\right)
\right) =
\]
\[
\frac 1{\left( 2\pi \hbar \right) ^{2\upsilon }}\int \int
dp_1dq_1dp_2dq_2\frac 12\left\{ F\left( p_1,q_1\right) +F\left(
p_2,q_2\right) \right\} \times
\]
\[
\times W\left( p_{1,}q_1;p_{2,}q_2;t;i\hbar \beta \right)
\]
where $\rho \left( 0\right) $ is the initial density matrix and $\tilde
Z=Tr\left( \rho \left( 0\right) \right) $. To obtain the explicit expression
of the terms of series (\ref{r3}) and to analyze its mathematical structure
let us introduce the pieces of dynamic trajectories defined by (\ref{a25})
according to the following recurrent relations:

\begin{equation}
\begin{array}{c}
\bar{p}_{j-1}^j=\bar{p}(\tau _{j-1};\bar{p}_j^{j+1}-s_j,\bar{q}_j^{j+1},\tau
_j) \\
\bar{q}_{j-1}^j=\bar{q}(\tau _{j-1};\bar{p}_j^{j+1}-s_j,\bar{q}_j^{j+1},\tau
_j) \\
\tilde{p}_{j-1}^j=\tilde{p}(\tau _{j-1};\tilde{p}_j^{j+1}-\eta _j,\tilde{q}%
_j^{j+1},\tau _j) \\
\tilde{q}_{j-1}^j=\tilde{q}(\tau _{j-1};\tilde{p}_j^{j+1}-\eta _j,\tilde{q}%
_j^{j+1},\tau _j)
\end{array}
\label{a33}
\end{equation}
for {\ initial conditions at times $\tau _j$} satisfying the following
inequalities $\{t=\tau _j\geq ...\geq \tau _2\geq \tau _1\geq \tau _0=0,$ $%
j=0,1,...\}$. If $\tau _j=t$ we assume that $s_j=0,$ $\bar{p}_j^{j+1}=p_{1,}$
$\bar{q}_j^{j+1}=q_1$ and $\eta _j=0,$ $\tilde{p}_j^{j+1}=p_2,$ $\tilde{q}%
_j^{j+1}=q_2$ and superscript in the left part of the expression is omitted $%
\bar{p}_{j-1}^j=\bar{p}_{j-1}$, $\bar{q}_{j-1}^j=\bar{q}_{j-1}$ and $\tilde{p%
}_{j-1}^j=\bar{p}_{j-1}$, $\tilde{q}_{j-1}^j=\bar{q}_{j-1}$. So for example
the third term $\left( j=3\right) $ can be written as:
\[
K_{\tau _2}^tK_{\tau _1}^{\tau _2}\bar{W}^{\tau _1}=\int_0^td\tau
_2\int_0^{\tau _2}d\tau _1\int ds_2d\eta _2\int ds_1d\eta _1\times
\]
\[
\times \gamma (s_2,\bar{q}_2;\eta _2,\tilde{q}_2)\gamma (s_1,\bar{q}%
_1^2;\eta _1,\tilde{q}_1^2)\bar{W}(\bar{p}_0^1,\bar{q}_0^1;\tilde{p}_0^1,%
\tilde{q}_0^1;i\hbar \beta )
\]

\section{Quantum dynamics}

The possibility to convert series like (\ref{a31}) into the form convenient
for probabilistic interpretation allow us to develop the Monte Carlo method
for its calculation \cite{filmd1} , \cite{filmd2} . According to the general
theory of the Monte Carlo methods for solving linear integral equations one
can simultaneously calculate all terms in the iteration series like (\ref
{a31}) \cite{sobol} . So using the basic ideas of \cite{sobol} we have
developed the Monte Carlo scheme , which provides domain sampling of the
terms giving the main contribution to the series (\ref{a31}). This sampling
reduces also numerical expenses for calculations of integrals of each terms
. Ensemble averaging on the configuration of interacting quantum particles
and classical heavy scatterers has been performed according to the
probability distribution proportional to $\left| \Psi \right| $ (\ref{a11}),
while the dynamic evolution has been realized according to equations (\ref
{a33}).

\section{System of units.}

Numerical calculations are more convenient to perform for dimensionless
equations. Let the quantum system considered have the following
characteristic energy $E_0$ and time $T_0$ scales. The dimensionless
combination $E_0T_0/\hbar $ may be considered as a measure of quantum
behaviour of a system. However in our quantum dynamics studies it is more
convenient to fix the maximum value of the considered time $t^{\prime }$ and
to take it as a unit of dynamic time $t$ of the system. So dimensionless
time $\tau =t/t^{\prime }$ will always vary from $0$ up to $1$. As unit of a
length we take the reciprocal wavenumber $k^{-1}$, determined by the ratio $%
k^2=2mE_0/\hbar ^2$, where $m$ is characteristic mass of quantum particle.
So, for example, for one electron in the external potential field $%
V_0U\left( q\right) $ the operator exponent of the time propagator can be
rewritten in the form:
\[
\hat{H}t/\hbar =\left\{ -\frac{\hbar ^2}{2m}\triangle +V_0U\left( q\right)
\right\} t/\hbar =
\]
\[
\left\{ -\frac{k^{-2}}{2M}\triangle +\frac{V_0}{2E_0M}U\left( q\right)
\right\} \tau =\left\{ -\frac 1{2M}\triangle +\xi _0U\left( q\right)
\right\} \tau
\]

Here $\triangle $ is Laplace operator, $V_0$ is characteristic constant of
interaction of potential field, $M= \hbar /2E_0t^{\prime }$, $\xi
_0=V_0/2E_0M=V_0t^{\prime }/ \hbar$, $\tilde t=t/t^{\prime }$. The similar
expressions and system of units were made use of for one electron in a field
of chaotic classical scatterers. In this system of units all the above
mentioned formulas take more simple form as $\hbar$ can be substituted for
by the unity, physical mass $m$ by formal parameter $M$ and constant of
interaction is redefined by relation $\xi _0=V_0t^{\prime }/ \hbar$.

A computation of the tensor of the electrical conductivity has been
performed as the first example of application of the developed approach.
According to quantum Kubo formulas in one electron approximation the tensor
of electrical conductivity is \cite{zubar} :
\[
\sigma _{\alpha \gamma }\left( \omega \right) =n\int_0^\infty \exp \left(
i\omega t-\epsilon t\right) \int_0^\beta \left\langle \hat{J}_\gamma \hat{J}%
_\alpha \left( t+i\hbar \lambda \right) \right\rangle d\lambda dt
\]
where $\epsilon \rightarrow 0$, $n$ is the electron density, $\hat{J}_\alpha
=e\dot{q}_\alpha \left( t\right) =ep_\alpha /m$ is the electrical current
operator , $\dot{q}_\alpha $ is the $\alpha $ component of electron velocity
operator. Wigner representation of this tensor may be written in the form:
\[
\begin{array}{c}
\sigma _{\alpha \gamma }\left( \omega \right) =n\int_0^\infty \exp \left(
i\omega t-\epsilon t\right) \int_0^\beta \phi _{\alpha \gamma }\left(
t,\lambda \right) d\lambda dt\equiv \\
\equiv nk^{-2}e^2\tilde{\sigma}_{\alpha \gamma }\left( \omega \right)
/2E_0t^{\prime } \\
\phi _{\alpha \gamma }\left( t,\lambda \right) =\left\langle \hat{J}_\gamma
\hat{J}_\alpha \left( t+i\hbar \lambda \right) \right\rangle = \\
=\frac 1{\left( 2\pi \hbar \right) ^{2\nu }}\int dp_1dq_1dp_2dq_2J_\gamma
\left( p_1,q_1\right) J_\alpha \left( p_2,q_2\right)
W(p_1,q_1;p_2,q_2;t;i\hbar \beta ;i\hbar \lambda )
\end{array}
\]
\[
\begin{array}{c}
W(p_1,q_1;p_2,q_2;t;i\hbar \beta ;i\hbar \lambda )=\int d\xi _1d\xi _2\exp
\left( i\frac{p_1\xi _1}\hbar \right) \exp \left( i\frac{p_2\xi _2}\hbar
\right) \times \\
\left\langle q_1+\frac{\xi _1}2\left| \exp \left( i\frac{\tau _1H}\hbar
\right) \right| q_2-\frac{\xi _2}2\right\rangle \left\langle q_2+\frac{\xi _2%
}2\left| \exp \left( -i\frac{\tau _2H}\hbar \right) \right| q_1-\frac{\xi _1}%
2\right\rangle
\end{array}
\]
where $\tau _1=t_1+i\hbar \lambda ,$, $\tau _2=t_2-i\hbar \left( \beta
-\lambda \right) $.
Here $\tilde{\sigma}_{\alpha \gamma }\left( \omega \right) $ is the
dimensionless tensor of electrical conductivity. In the model of independent
electrons in the medium of chaotic classical scatterers (one particle
approximation) the Hamiltonian of the system can be written in the form:
\[
\hat{H}=\hat{H}_{es}+\tilde{H}_{ss}^{cl}
\]
\[
\hat{H}_{es}=-\triangle /2M+\sum_{j=1}^N\xi _0^{es}U_{es}\left( \left|
q-Q_j\right| /\sigma ^{\prime }\right)
\]
\[
\tilde{H}_{ss}^{cl}=\sum_{j=1}^Np^2m/2Mm_s+\sum_{i\neq j}^N\xi
_0^{ss}U_{ss}\left( \left| Q_i-Q_j\right| /\sigma ^{\prime \prime }\right)
\]
where $m_s$ is the mass of scatterer $\left( m/m_s\ll 1\right) $, $\tilde{H}%
_{ss}^{cl}$ is the classical Hamiltonian of scatterers, $\xi
_0^{es}=V_0^{es}t^{\prime }/\hbar $ and $\xi _0^{ss}=V_0^{ss}t^{\prime
}/\hbar $ are constant of interaction of pair vise electron-scatterer $%
U_{es}\left( \left| q-Q_j\right| /\sigma ^{\prime }\right) $ and
scatterer-scatterer $U_{ss}\left( \left| Q_i-Q_j\right| /\sigma ^{\prime
\prime }\right) $ potentials respectively, $Q_j$ are the scatterers
positions ($j=1,..N$), $\sigma $ is the characteristic length of potential.

Ensemble averaging on the configuration of interacting electron and
classical heavy scatterers in quantum Kubo formulas has been performed
according to the probability distribution proportional to $\left| \Psi
\right| $ , while the dynamic evolution has been realized according to
recurrent relations (\ref{a33})

The potential barrier of real type for electron -scatterer and
scatterer-scatterer interaction has been taken in the Gaussian form:
\begin{equation}  \label{a74}
U\left( \left| q-q^{\prime }\right| /\sigma \right) =\exp \left( -\left|
q-q^{\prime }\right| ^2/\sigma ^2\right)
\end{equation}
with equal to each other the all constants of interaction ( $\xi _0^{es}=\xi
_0^{ss}$ $>0$ and $\sigma ^{\prime }=\sigma ^{\prime \prime }$).

\section{Numerical results}

Anderson localization of electrons have been investigated for one, two- and
three dimensional disordered systems of scatterers.
For 1D case the results of our simulation and their brief description are
presented below. In all calculations the dimesionless density of classical
scatterer is approximately equal to unity $\left( n\sigma \approx 1\right) $%
, while $k\sigma \approx 1$ and $k^{-1}n\approx 1$.

\subsection{Electron conductivity of disordered systems of scatterers}

For the two small ratios of temperature to the height of a potential barrier
($kT/V_0$) equal to 0.04 and 0.0025 the Fig. \ref{fig:crrchd1} and Fig. \ref
{fig:crqchd1} present 1D results for diagonal elements of the tensor of the
electrical conductivity, which are the momentum- momentum time correlation
functions. Curves 1 on both figures relate to calculations taking into
account only one term of iteration series (\ref{a31}), while curve 2 present
results allowing for all terms of iteration series (\ref{a31}). So curves 1
on both figures have been obtained by using only the classical trajectories
without momentum jumps. The curve 1 on the Fig. \ref{fig:crrchd1} have
practically the traditional fast decay.

Curves 2 on both figures present momentum- momentum time correlation
function obtained for dynamic trajectories with momentum jumps. The
difference between these two correlation functions is larger at lower
temperatures. Analyzing Fig. \ref{fig:crqchd1} one can conclude that curves
1 and 2 show undamped oscillations. Appearing of the undamped oscillations
at very low temperature results from shortcomings of our model, in which the
classical slow heavy scatterers are treated as non- moving particles on the
time scale of the order of characteristic dynamic time of electrons.

However as is known from the literature the undamped oscillations results in
unphysical behaviour of the Fourier transform of the momentum- momentum time
correlation function. The next Fig. \ref{fig:spcfm1d}, \ref{fig:spqcm1d},
\ref{fig:spichd1} and \ref{fig:spiqchd1} show the significant difference of
the Fourier transforms for correlation function with damped and undamped
oscillations. 
Fig. \ref{fig:spcfm1d}, \ref{fig:spqcm1d} and \ref{fig:spichd1}, \ref
{fig:spiqchd1} demonstrate the real and imaginary parts of its Fourier
transforms versus the dimensionless frequency $\hbar \omega /V_0$. The real
part of Fourier transform characterizes the Ohmic absorption of
electromagnetic energy and has the physical meaning of electron
conductivity, while the imaginary part presents $(\epsilon -1)\omega $,
where $\epsilon $ is permittivity of the system. The curve 2 (quantum
trajectories) on Fig. \ref{fig:spcfm1d} is higher than the curve 1
(classical trajectories) but both curves are going to zero at small
frequency pointing out that the static conductivity at zero frequency is
equal to zero or is very small. Note that points of curve 2 are going to
zero faster than the same value obtained in approximation of classical
trajectories (curve 1). This behaviour of conductivity in the vicinity of
zero frequency is the characteristic manifestation of the electron Anderson
localization.

At lower temperatures the undamped oscillations with the proper phase shift
of the momentum- momentum time correlation function may result in appearance
of the negative values of real part of Fourier transform on Fig. \ref
{fig:spqcm1d}. To overcome this shortcomings of our model at very low
temperature it is necessary to take into account the slow motion of heavy
particles, which should destroy the coherence oscillation of the light
electrons trapped by heavy particles and cancel these negative values.

\subsection{Position and momentum dispersions}

The Fig. \ref{fig:dxchd1} presents the 1D results for quantum position
dispersion versus the dimesionless time $tV_0/ \hbar$ at the mentioned above
two ratios of temperature to the height of a potential barrier ($kT/V_0$).
Curves 1 and 2 correspond to $kT /V_0$ equal to 0.04 and 0.0025 respectively.

In one dimensional case quantum electrons is known \cite{leeram} should be
localized at any temperatures but the length of localization is function of
the particle energy. Analyzing data, submitted on the Fig. \ref{fig:dxchd1},
it is possible to note that at lower temperature the curve 2 is very flat
and the localization length ($\lambda $) can be estimated as the squared
root of the characteristic value of the position dispersion ($\lambda \leq
0.7$). At higher temperature the behaviour of curve 1 is more complicated.
At the initial stage ($tV_0/\hbar \leq 100$) curve 1 has a very flat part,
which can be also considered as a characteristic manifestation of
localization.

However when the time is more than 100 the curve 1 shows behaviour, which is
typical for particle diffusion. This characteristic change in behaviour may
be connected with momentum jumps of quantum trajectories. The virtual energy
of the trajectories due to momentum jumps may be large. So the exponentially
rare fast trajectories with very large value of localization length can give
the exponentially large contribution to position dispersion as the
difference in positions of fast trajectories may be exponentially large at
large enough time. The similar problem connected with the exponentially
large contribution of the exponentially rare events arises at the
consideration of the classical wave propagation in random media. It is known
\cite{gred1} for one dimensional case that the dispersion of wave intensity
is not self- averaging value as the exponentially rare configurations of
scatterers can give the exponentially large contribution of intensity at
large enough distances from the wave source. Assuming that the initial flat
part of curve 2 indicates the particle localization we can estimate the
localization length as equal to 1.2.

The Fig. \ref{fig:dpchd1} presents momentum dispersion of quantum particle
in disordered systems of scatterers. Curves 1 and 2 are stabilized and it is
interesting to note that the lower temperature curve 2 are stabilized at
larger value than curve 1. This may be connected with the uncertainty
momentum- position principle and related fast oscillation of the localized
quantum particles.

\subsection{Energy distribution function}

Quantum average energy of the system has been calculated according to the
easily derived formula:
\[
\bar H\left( t\right) =\frac 1{\left( 2\pi h\right) ^{2\upsilon }}\int
dp_1dq_1dp_2dq_2\frac 12\left\{ H\left( p_1,q_1\right) +H\left(
p_2,q_2\right) \right\} \times
\]
\[
W\left( p_1,q_1;p_2,q_2;t;ih\beta \right)
\]

Our calculations have shown that this function versus time variable is
practically constant. To analyze the energy fluctuations during quantum
dynamic evolution of the system the more interesting energy distribution
function has been also calculated. The simple semi classical approximation
of this function is defined by the following expression:
\[
\rho \left( E,\beta \right) =
\]
\[
\frac 1{\left( 2\pi h\right) ^{2\upsilon }}\int_0^1dt\int
dp_1dq_1dp_2dq_2\frac 12\left\{ \delta \left( E-H\left( p_1,q_1\right)
\right) +\delta \left( E-H\left( p_2,q_2\right) \right) \right\} \times
\]
\[
W\left( p_1,q_1;p_2,q_2;t;ih\beta \right)
\]

The function $\ln \left( \rho \left( E,\beta \right) \right) $ versus $E/V_0$
has been presented by curves 1 and 2 on the Fig. \ref{fig:ed10f1d} and \ref
{fig:ed251d} for higher and lower temperatures respectively. Curves 1
present results obtained in approximation of classical trajectories (only
first term in iteration series), while curves 2 show results allowing for
all terms of iteration series. At $E/V_0\leq 4$ curves 1 and 2 practically
coincide both at higher and lower temperatures. However at the $E/V_0\geq 4$
the distinction in curves 1 and 2 is large due to momentum jumps resulting
in appearance of trajectories with large virtual energy. So the asymptotic
behaviour of the curves 1 and 2 is quite different. These asymptotics in
logarithmic scale may be fitted by the strait lines with different slope.
The existence of the long exponentially decreasing tail of the energy
distribution function supports our explanation of the peculiarity in the
behaviour of position dispersion of quantum particles. For detailed
investigations the additional consideration is needed to analyze this effect.

Lowering temperatures results also in spliting the main peak in energy
distribution function into two sharper peaks. As it follows from Fig. \ref
{fig:ed10f1d} and \ref{fig:ed251d} the energy gap at the edge of the energy
spectrum a bit increases with the lowering temperature.

\section{Conclusion}

In the Wigner formulation of quantum statistical mechanics for canonical and
micro canonical ensembles we have presented a new computational technique
allowing quantum dynamics simulations for systems including subsystems of
quantum interacting particles and subsystems of classical heavy scatterers
as well as the system of quantum particles in external potential field. The
developed approach for quantum dynamics includes a sophisticated combination
of well known molecular dynamics method and Monte Carlo technique.

For electrons in disordered systems of scatterers the numerical results have
been obtained for series of the average values of the quantum operators
including position and momentum dispersions, average energy, energy
distribution function as well as for the frequency dependencies of tensor of
electron conductivity and permittivity according to quantum Kubo formula.
Zero or very small value of static conductivity have been considered as the
manifestation of Anderson localization of electrons in 1D case. Independent
evidence of Anderson localization comes from the behaviour of the calculated
time dependence of position dispersion. Nevertheless for localized electrons
the energy distribution function obtained has the long exponentially
decaying tail, which is the reason of exponentially rare appearance of large
values of quantum particle virtual energy that strongly affects the behaviour
of the position dispersion.

\section{Acknowledgments}

The authors are very appreciated to Professor K. Singer for fruitful
discussions, invaluable comments and interest to work. The authors expresses
thanks to Russian Fund for Basic Researches for financial support of this
work ( grants 97- 02- 16572, 97-1-00931, 96-1596462 ).


\newpage

\begin{figure}
\caption[Mmtcrfl]
   { \label{fig:crrchd1}
Momentum-momentum time correlation function versus $tV_0/ \hbar$
for higher temperature ($kT /V_0 = 0.04$):
1- approximation of the classical trajectories;
2- quantum trajectories}
\caption[Mmtcrfh]
   { \label{fig:crqchd1}
Momentum-momentum time correlation function versus $tV_0 / \hbar$
for lower temperature ($kT /V_0 = 0.0025$):
1- approximation of the classical trajectories;
2- quantum trajectories}
\end{figure}

\begin{figure}
\caption[sprh]
   { \label{fig:spcfm1d}
The real part of Fourie transform of the momentum-momentum time correlation
function versus $\hbar \omega /V_0 $ for higher temperature ($kT /V_0=0.04$):
1- approximation of the classical trajectories;
2- quantum trajectories}
\caption[sprl]
   { \label{fig:spqcm1d}
The real part of Fourie transform the momentum-momentum time correlation
function versus $\hbar \omega /V_0 $ for lower temperature ($kT /V_0$=0.0025):
1- approximation of the classical trajectories;
2- quantum trajectories}
\end{figure}

\begin{figure}
\caption[spih]
   { \label{fig:spichd1}
The imaginary part of Fourie transform of the momentum-momentum time correlation
function versus $\hbar \omega /V_0 $ for higher temperature ($kT /V_0=0.04$):
1- approximation of the classical trajectories;
2- quantum trajectories}
\caption[spil]
   { \label{fig:spiqchd1}
The imaginary part of Fourie transform the momentum-momentum time correlation
function versus $\hbar \omega /V_0 $ for lower temperature ($kT /V_0$=0.0025):
1- approximation of the classical trajectories;
2- quantum trajectories}
\end{figure}

\begin{figure}
\caption[Positiondisp]
   { \label{fig:dxchd1}
Position dispersion: 1- higher temperature ($ kT \omega /V_0 = 0.04 $);
2- lower temperature ($kT /V_0 = 0.0025$)}
\caption[Momentum disp]
   { \label{fig:dpchd1}
Momentum dispersion: 1- higher temperature ($kT /V_0 = 0.04$);
2- lower temperature ($kT /V_0 = 0.0025$)}
\end{figure}

\begin{figure}
\caption[Momentum]
   { \label{fig:ed10f1d}
Energy distribution in logarithmic scale versus $E_0/V_0$ for higher temperature
($kT /V_0=0.04$):
1- approximation of the classical trajectories;
2- quantum trajectories}
\caption[Momentum]
   { \label{fig:ed251d}
Energy distribution in logarithmic scale versus $E_0/V_0$ for lower temperature
($kT /V_0=0.0025$):
1- approximation of the classical trajectories;
2- quantum trajectories}
\end{figure}





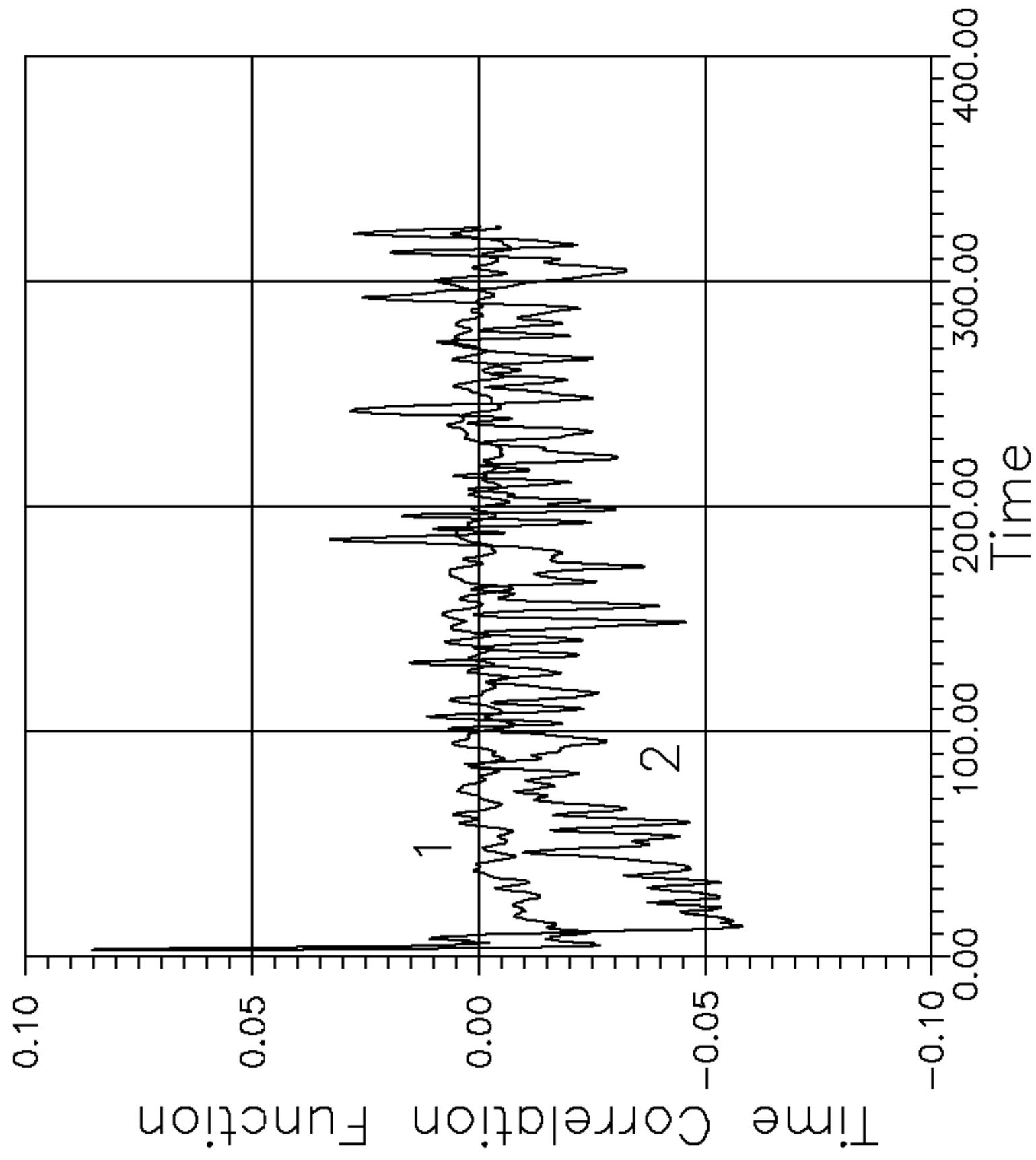
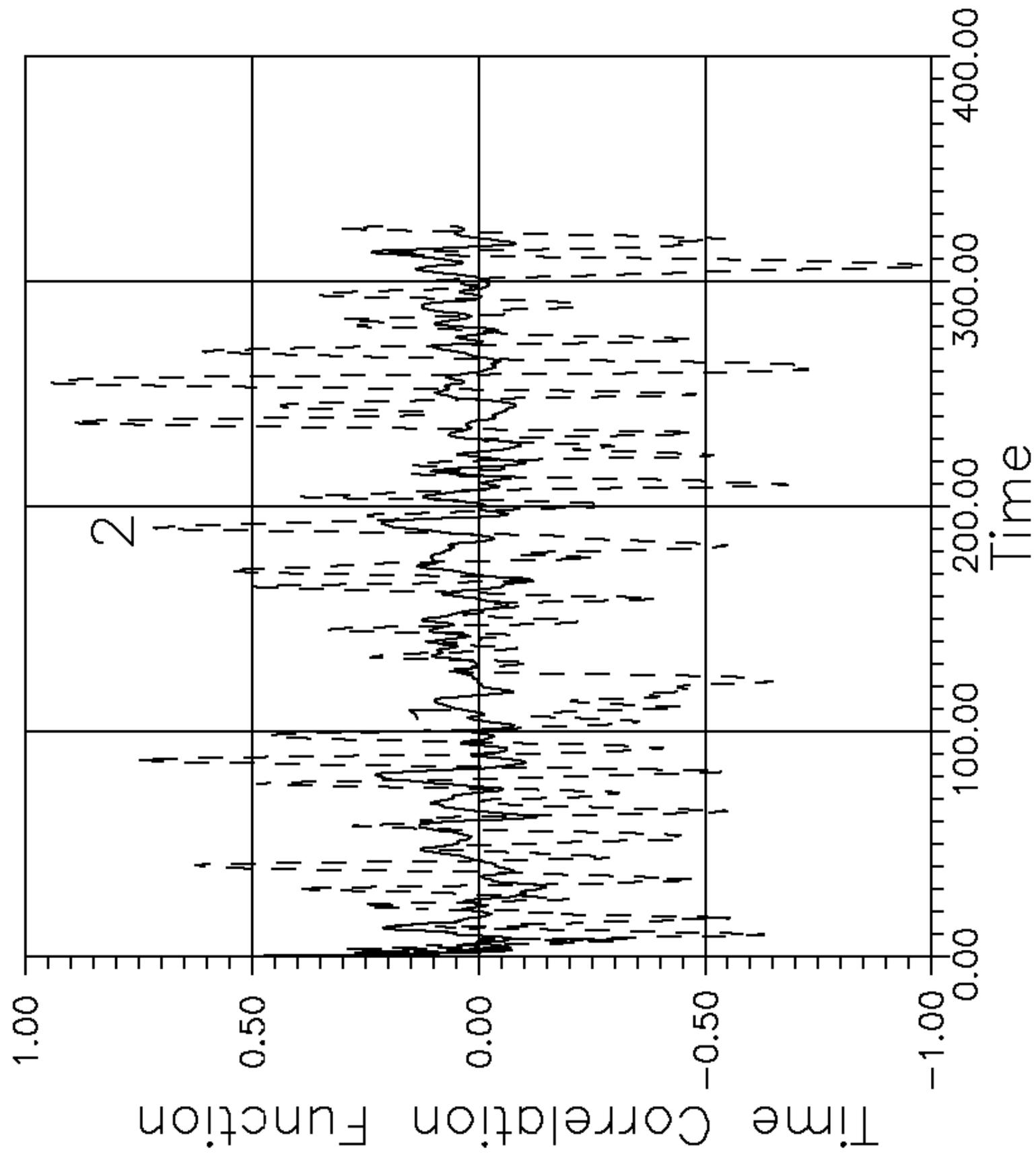
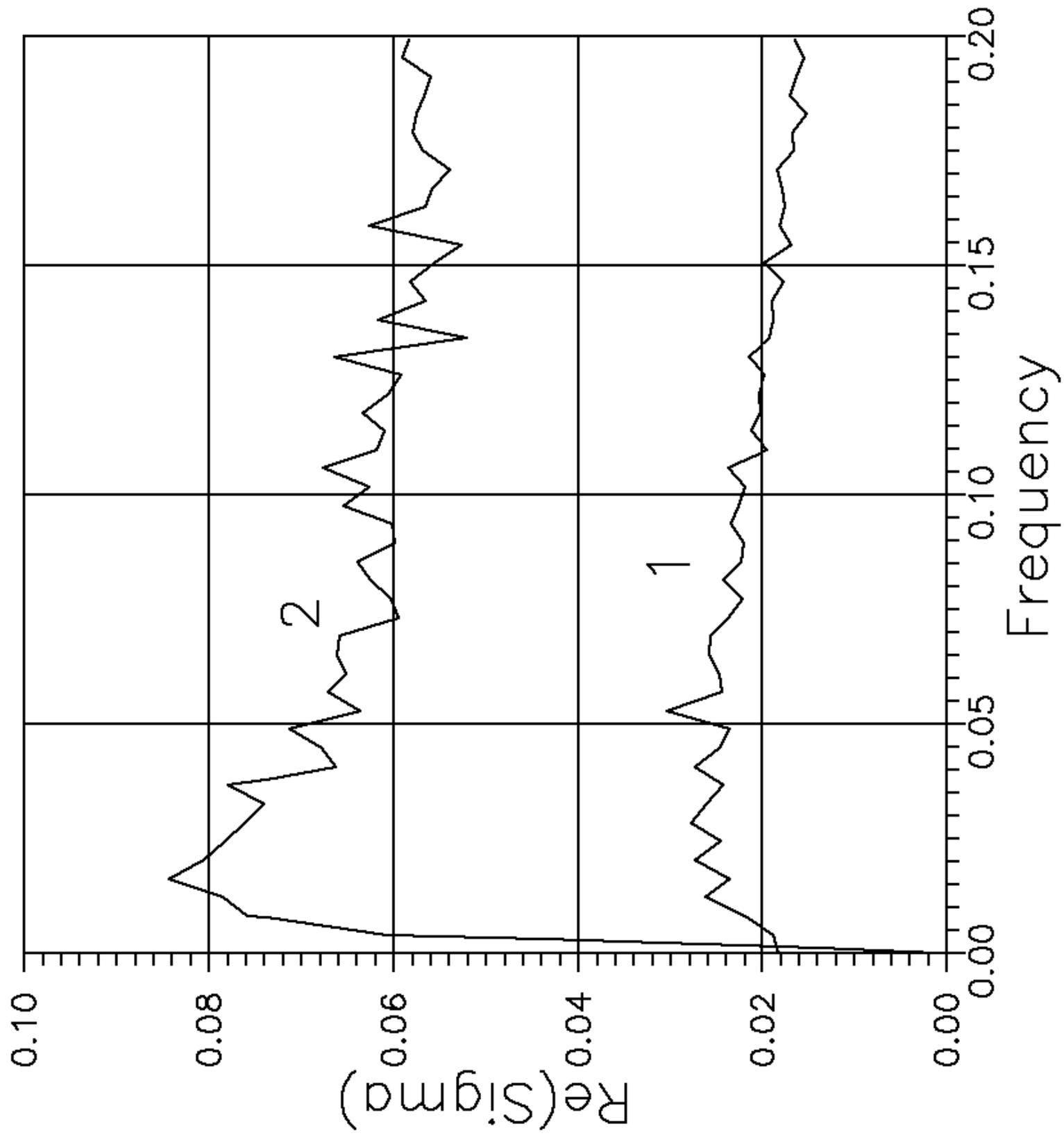
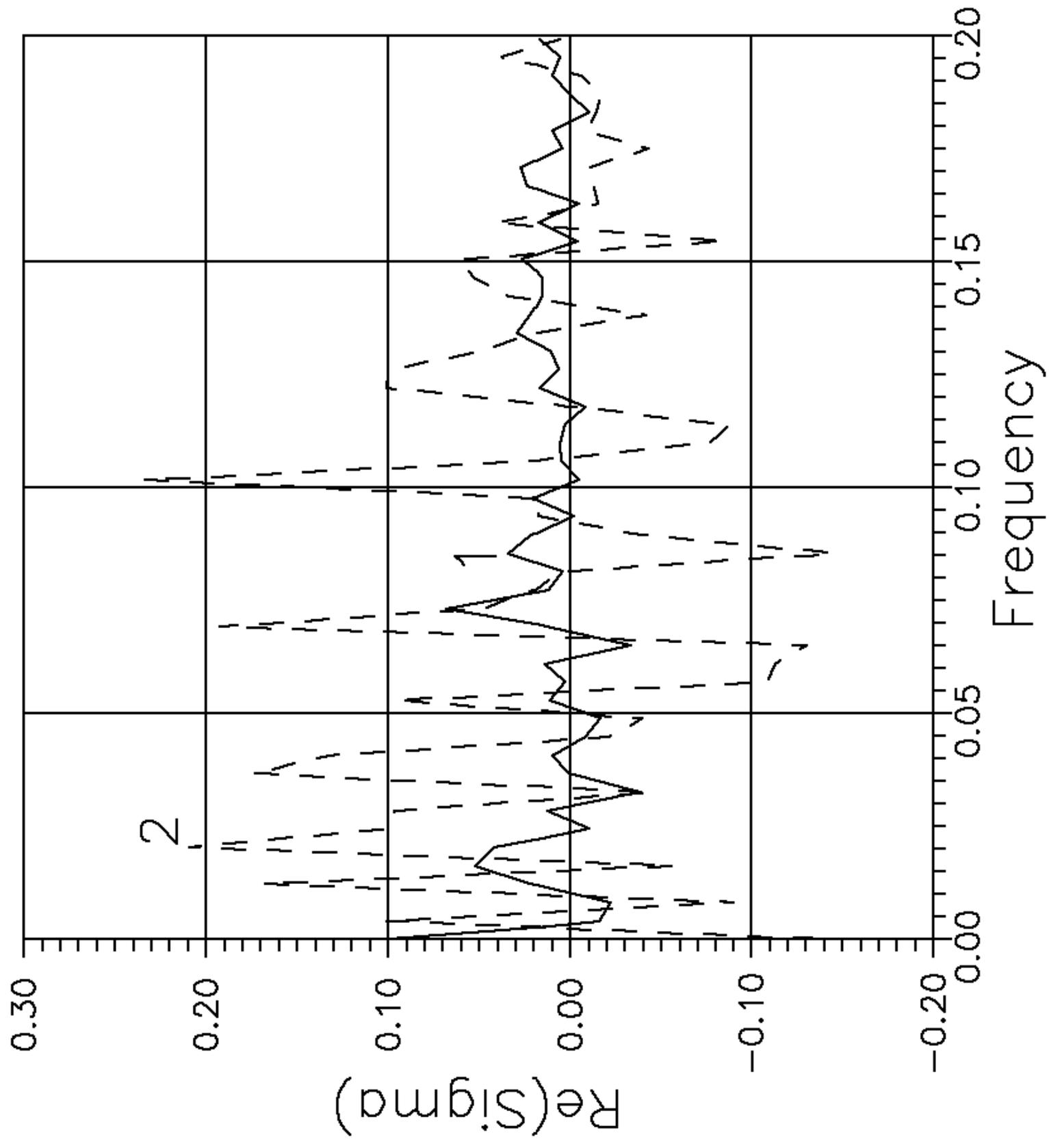
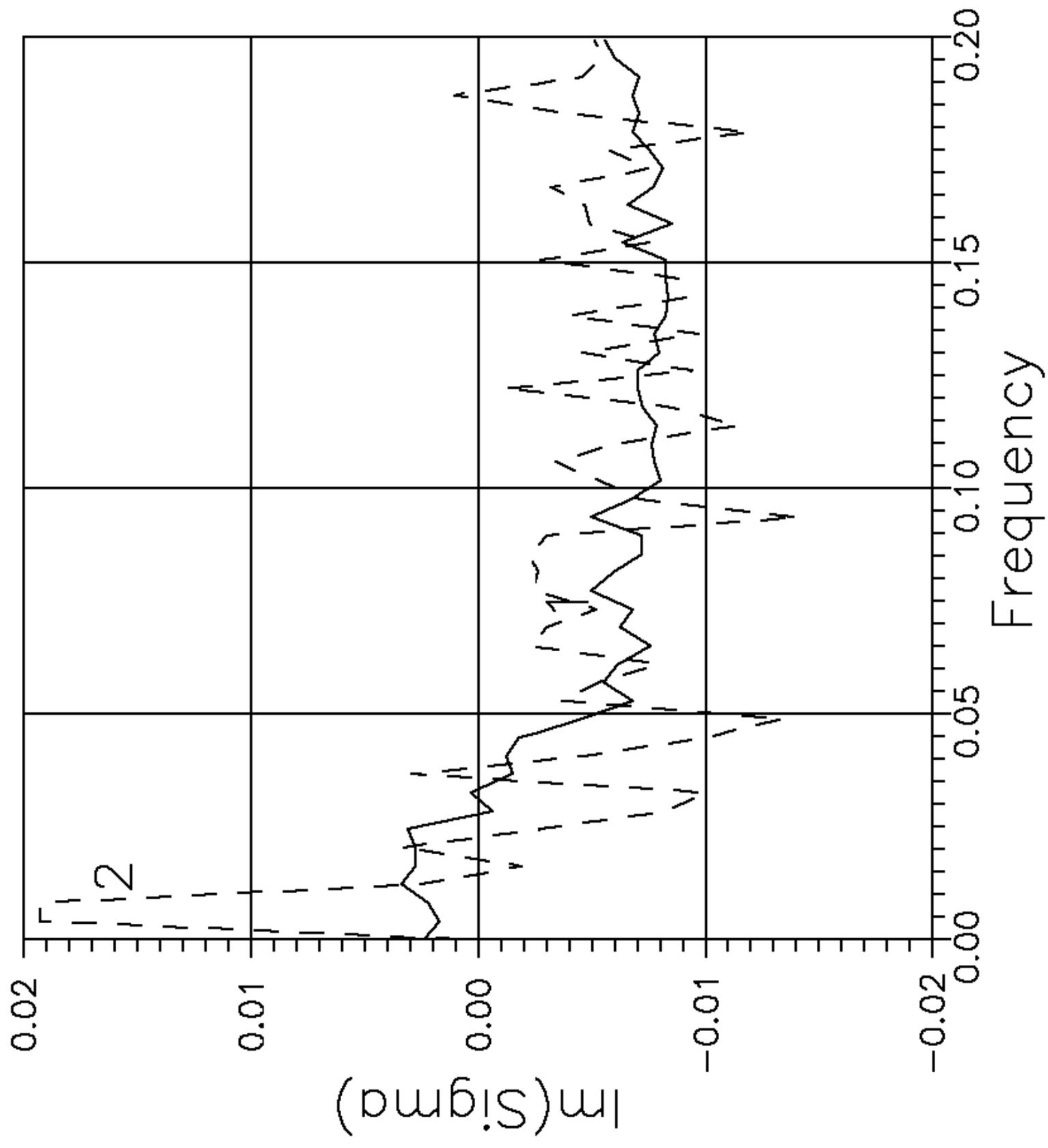
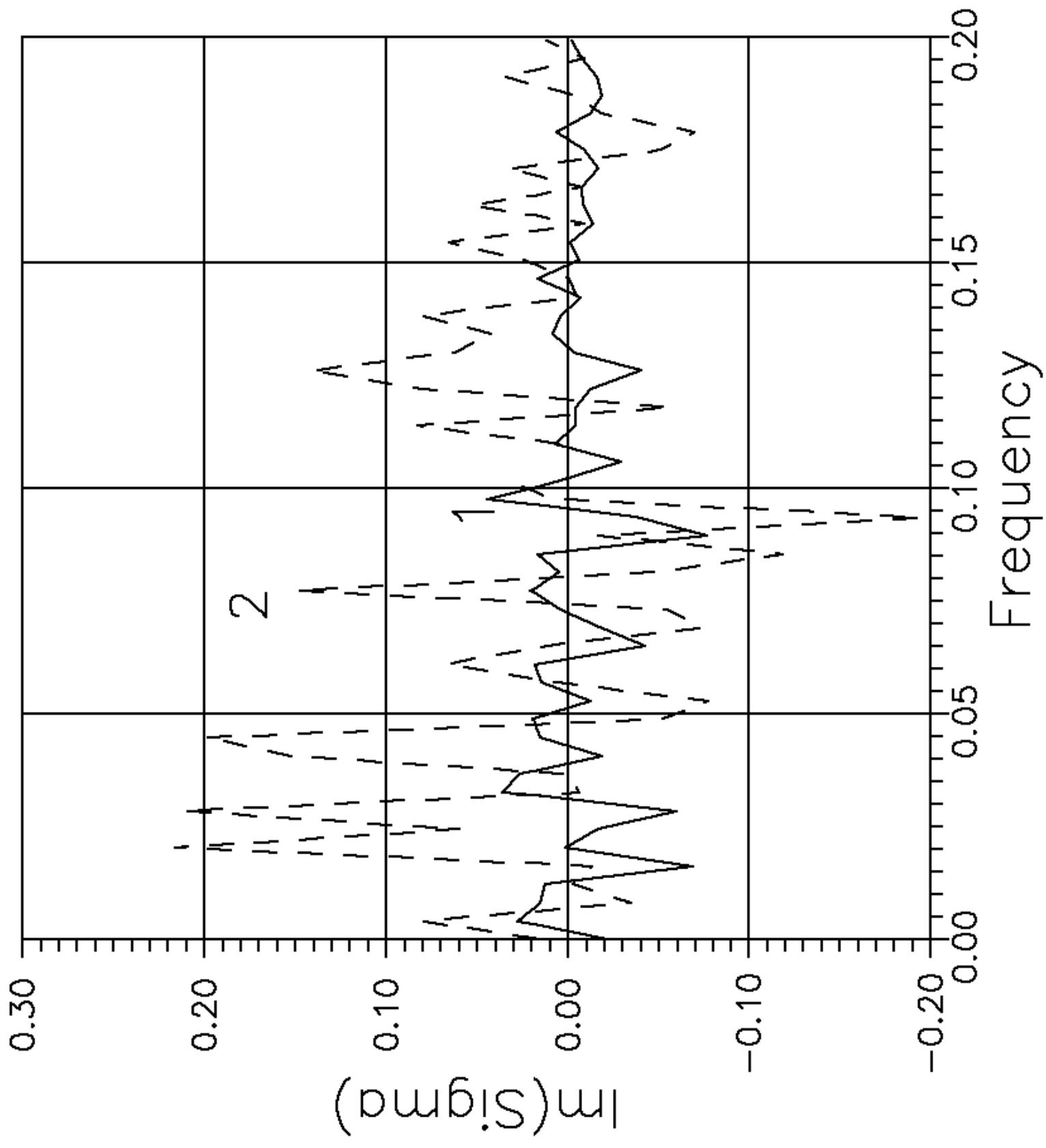
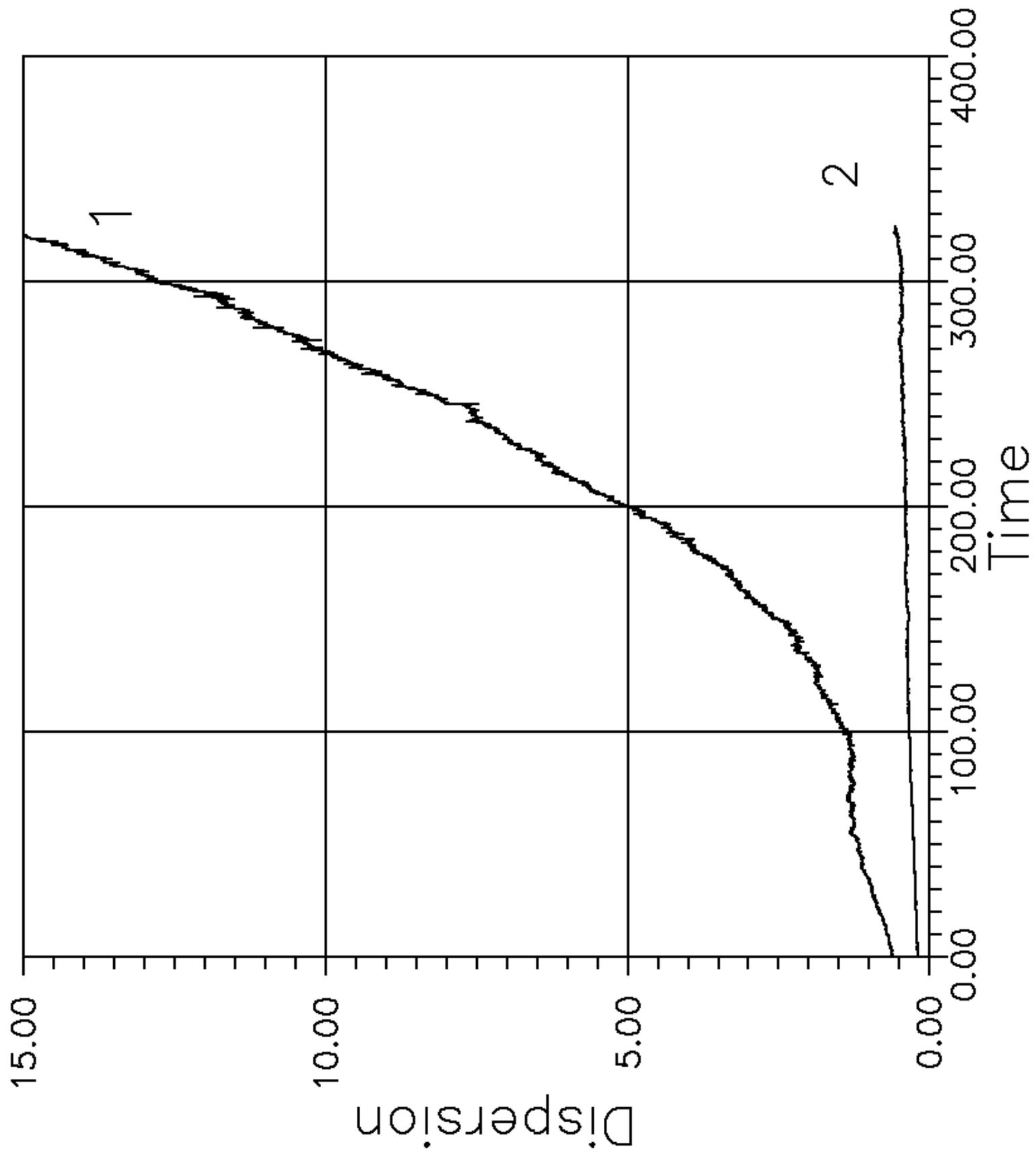
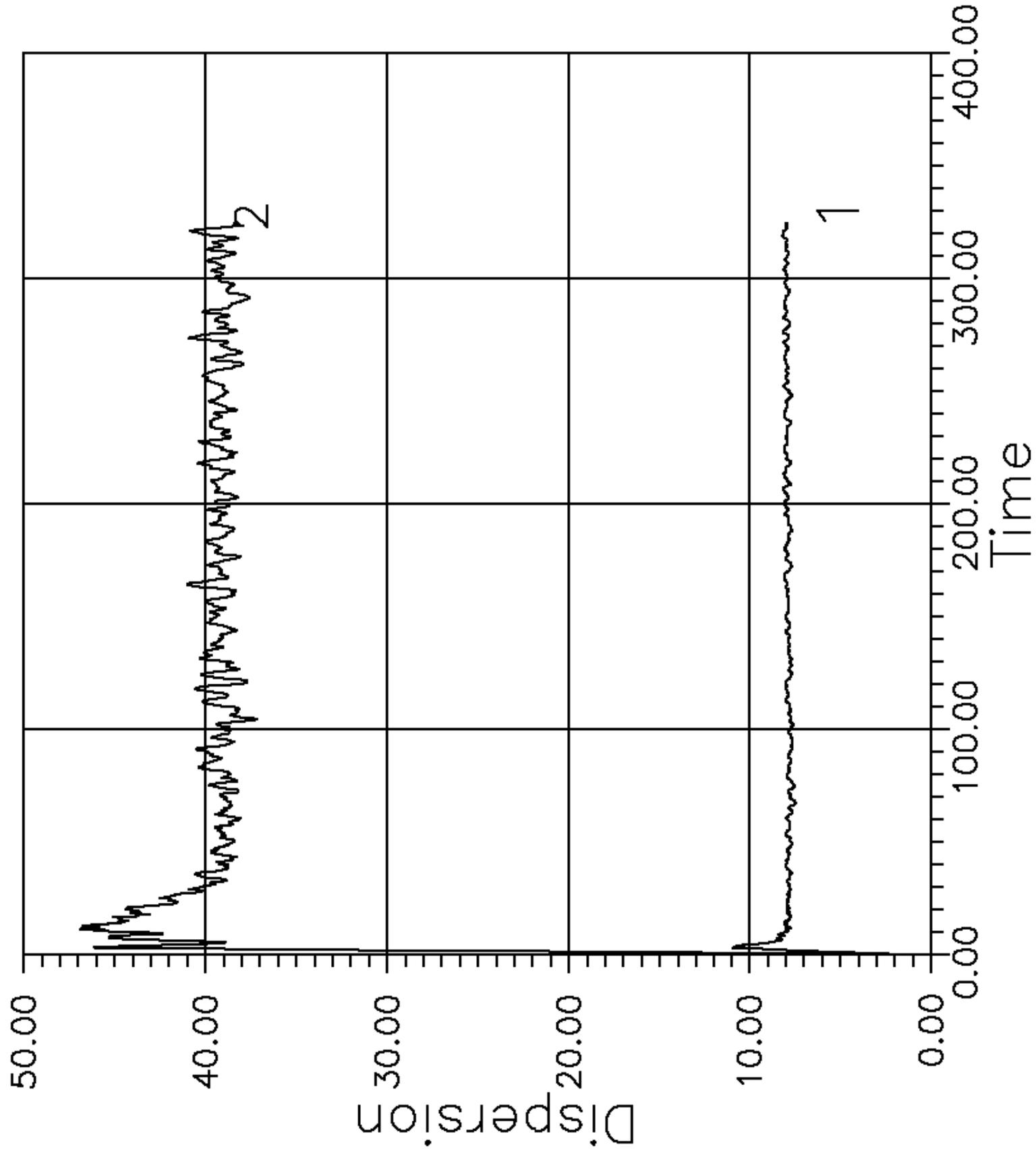
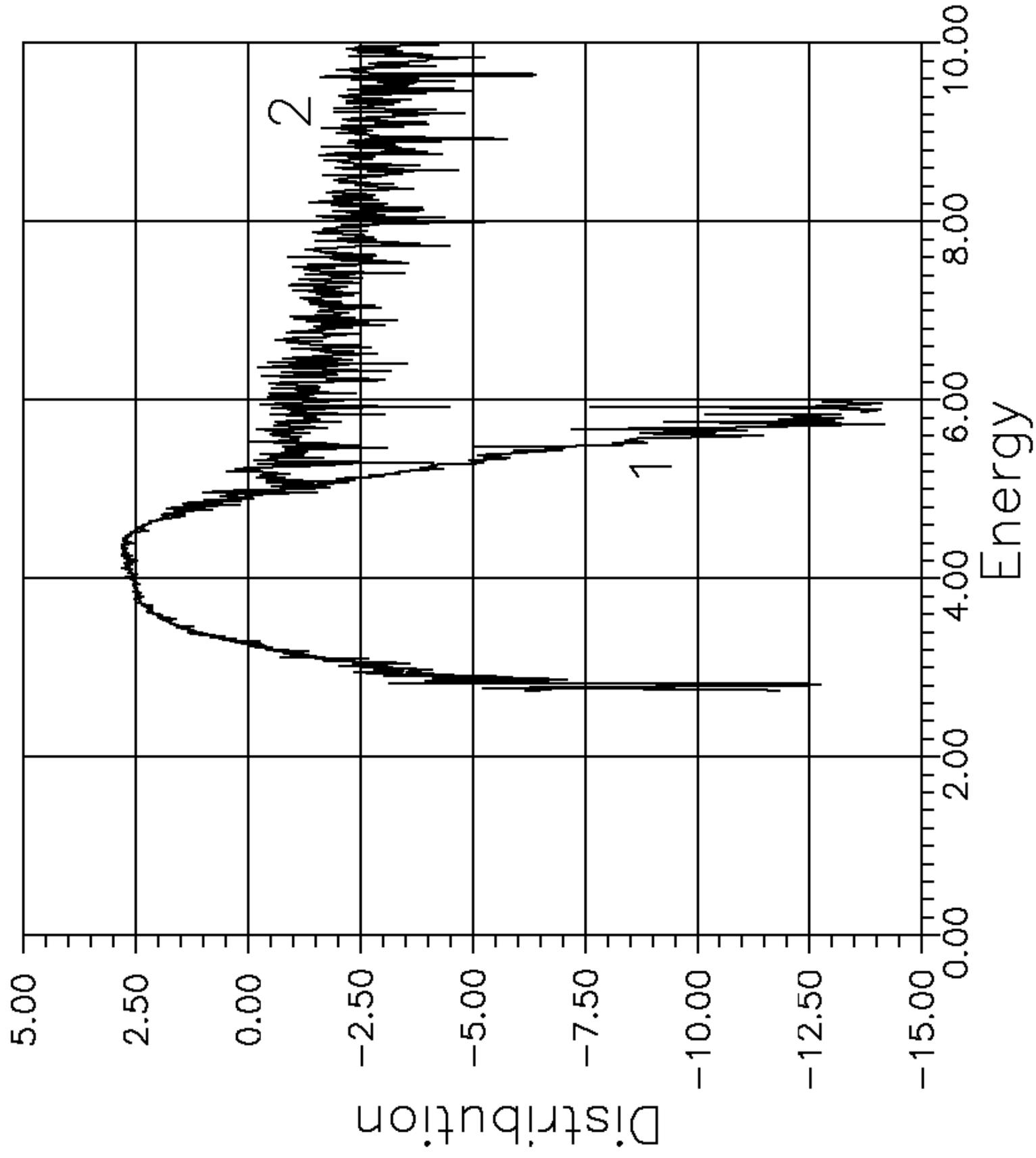
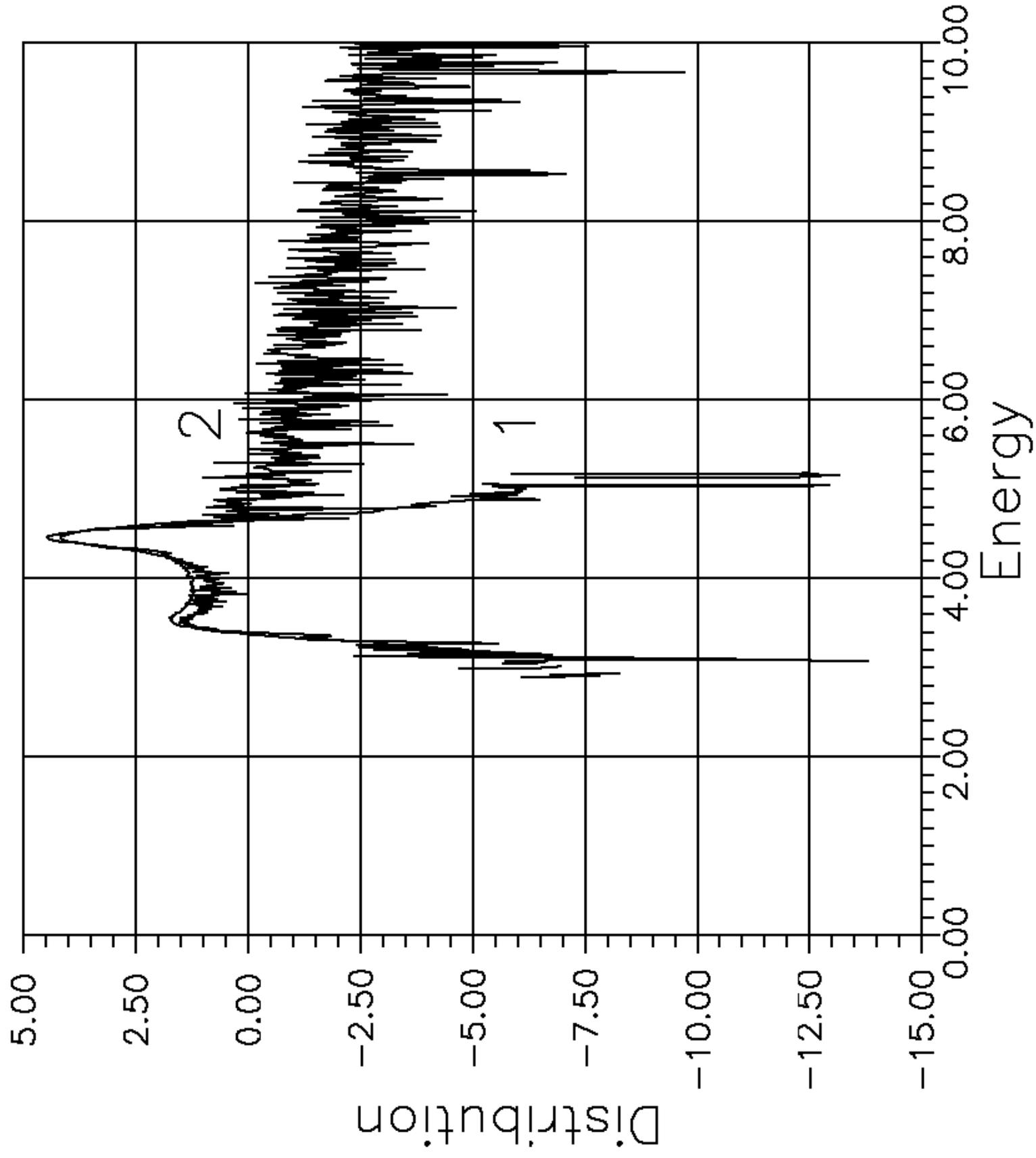
\end{document}